\documentstyle[emulateapj,epsfig]{article}

\begin{document}
\submitted{Accepted for publication in ApJ Letters}

\title{
The Formation of the First Globular Clusters in
Dwarf Galaxies Before the Epoch of Reionization} 

\author{Volker Bromm$^{1}$ and Cathie J. Clarke}
\affil{Institute of Astronomy, University
of Cambridge, Madingley Road, Cambridge CB3 0HA, UK;\\
volker@ast.cam.ac.uk, cclarke@ast.cam.ac.uk}

\begin{abstract}
We explore a mechanism for the formation of the first globular clusters,
operating during the assembly of dwarf galaxies at high redshifts, $z\ga 10$.
The substructure in the dark matter and the corresponding potential wells
are responsible for setting the cluster scale of $\sim 10^{5}M_{\odot}$.
The second mass scale in the formation problem, the stellar scale of
$\sim 1 M_{\odot}$, is determined in turn by the processes that cool the gas.
We address the origin of the first, cluster mass scale by means of three-dimensional
numerical simulations of the collapsing dark matter and gaseous components.
We find that the gas falls into the deepest dark subhalos, resulting in a
system of $\sim 5$ proto-globular clouds. The incipient globular clusters
lose their individual dark halos in the process of violent relaxation, leading
to the build-up of the general dark halo around the dwarf galaxy.
%This process removes the prime objection to the dark matter induced
%formation of globular clusters.

\end{abstract}

\footnotetext[1]{Present address: Harvard-Smithsonian Center for 
Astrophysics, 60 Garden Street, Cambridge, MA 02138}

\section{INTRODUCTION}

The origin of globular clusters (GCs) is a longstanding challenge
in astrophysics,
ever since Peebles \& Dicke (1968) have tried to link their formation
to the conditions in the early universe, briefly after the epoch of
recombination (at redshift $z\sim 1000$). Despite considerable progress
in our theoretical understanding (e.g., Fall \& Rees 1985; Ashman 1990;
Kang et al. 1990; Murray \& Lin 1992;
McLaughlin \& Pudritz 1996;
Padoan, Jimenez, \& Jones 1997; Nakasato, Mori, \& Nomoto 
2000; Ashman \& Zepf 2001; Cen 2001), 
the fundamental question of why it is that at early cosmological
times bound aggregates of $\sim 10^{5}$ stars were able to form, remains
unsolved. Two recent developments, however, have significantly improved
the prospects for real progress.

Observations with the {\it Hubble Space Telescope} have revealed several
merging or recently merged systems which contain extremely luminous
young stellar aggregates (e.g., Whitmore \& Schweizer 
1995). These systems have estimated
sizes and masses close to those of Galactic GCs. Therefore, in extreme
environments like starburst galaxies, GCs might still be able to form
in the present-day universe where we can directly probe the formation
process.
The second recent development is the increase in
computational power which enables us to address the complex
physics of collapsing and fragmenting gas with high resolution and
the addition of the important physical processes.
Therefore, the renewed investigation of globular cluster formation
is very timely, and holds the promise of understanding star
formation on its grandest scale.

It is likely that there are many avenues leading to the formation
of GCs (e.g., Ashman \& Zepf 1998). In this {\it Letter}, we explore
one of them and ask: {\it Could the first globular clusters have formed
during the initial stages in the hierarchical build-up of cosmic 
structure?} Within popular variants of the `cold dark matter' (CDM) model,
the first dwarf galaxies of mass $\sim 10^{8}M_{\odot}$ are expected
to collapse at $z \ga 10$. Forming GCs in these dwarf systems might
provide an answer to the old puzzle of how to simultaneously account
for the two characteristic mass scales involved in the
problem: the cluster scale of $\sim 10^{5}
%problem (e.g., Zinnecker \& Palla 1987): the cluster scale of $\sim 10^{6}
M_{\odot}$, and the stellar scale of $\sim 1M_{\odot}$, respectively.
In the context of our model, the cluster scale is set by the substructure
in the dark matter (DM) component, providing the potential wells in which
the proto-globular clouds are assembled.
The second, stellar, mass scale is then determined by the cooling
physics of the star forming gas. We here address the origin of the cluster
mass scale by performing numerical simulations of the collapsing DM and
gas components. The emergence of the stellar mass scale will be investigated
in subsequent, higher-resolution work.

The possible connection between GC formation and DM subhalos has
previously been explored semi-analytically by Peebles (1984), Rosenblatt,
Faber, \& Blumenthal (1988), and C\^{o}t\'{e}, West, \& Marzke (2001).
Following the demonstration that GCs do not have individual dark
halos (Moore 1996), this scenario had lost much of its initial appeal.
Considering GC formation during the early stages of the CDM bottom-up
hierarchy, however, might also provide a way out of this problem. Due
to the flatness of the CDM spectrum on the smallest scales, there is
a strong `cross-talk' behavior,i.e., the roughly simultaneous collapse
of all scales (e.g., Blumenthal et al. 1984). The incipient GCs, provided they
have condensed sufficiently, could therefore lose their
individual dark halo in the process of violent relaxation without being 
disrupted themselves.

\section{THE COSMOLOGICAL CONTEXT}

We first outline the basic physical reason why small protogalactic
systems of mass $\sim 10^{8}M_{\odot}$, collapsing before the epoch
of reionization at redshifts $z\ga 7$ (e.g., Gnedin 2000),
provide intriguing sites for the formation of the first globular clusters.
Throughout this paper, we assume that structure formation is described
by the $\Lambda$CDM model, with a density parameter in matter (both
DM and gas) of
$\Omega_{m}=1-
\Omega_{\Lambda}=0.3$, and in baryons of
$\Omega_{\rm B}=0.045$.
The Hubble constant is $h=H_{0}/(100$ km s$^{-1}$ Mpc$^{-1}$)=0.7, and
the power-spectrum normalization is given by
$\sigma_{8}=0.9$ on the $8h^{-1}$Mpc scale. 
%%(see Barkana \& Loeb 2001).

Within a hierarchical cosmogony, the
first dwarf galaxies, of mass $\sim 10^{8}M_{\odot}$, are expected to form out
of $\sim 3 \sigma$ peaks in the random field of primordial density fluctuations.
One can estimate the
redshift of collapse (or virialization) to be
$z_{coll}\simeq 15$.
For such a system, the gas acquires a temperature close to the virial
temperature of the DM host halo: $T_{vir}\simeq 2000\mbox{\,K}
h^{2/3}(M/10^{8}M_{\odot})^{2/3}(1+z_{coll})\simeq 10^{4}\mbox{\,K\ }$.
Since structure formation is a bottom-up process, at $z>z_{coll}$ less massive
DM subhalos with $T_{vir}<10^{4}\mbox{\,K}$ will collapse first, merging
to eventually build up the dwarf galaxy.
The $3 \sigma$ peaks considered here are very likely to be
incorporated into larger systems later on. The fate of dwarf
galaxies deriving from lower $\sigma$ peaks, however, is less
certain. Whereas $2 \sigma$ peaks are still expected to collapse
before reionization, the more typical $1 \sigma$ peaks would
collapse at lower redshift and would therefore not form GCs according
to our model. The $2 \sigma$ peaks could survive as individual entities
and thus correspond to Local Group dwarfs like Fornax and Sagittarius
which are observed to have a system of GCs.

A convenient way to quantify the merging
history of a galaxy is given by the extended Press-Schechter (EPS)
formalism (Lacey \& Cole 1993). The EPS prescription allows
one to compute the average number of progenitor subhalos at redshift
$z$ in a unit range of ln$M$ that will have merged at a later time
$z_{0}=15$ into a more massive halo of mass $M_{0}=2\times 10^{8}M_{\odot}$:
\begin{equation}
\frac{{\rm d}N}{{\rm d ln}M}=\sqrt{\frac{2}{\pi}}M_{0}\sigma_{0}(M)
\frac{D}{S^{3/2}}{\rm exp}\left(-\frac{D^{2}}{2S}\right)
\left|\frac{\rm{d}\sigma_{0}(M)}{\rm{d}M}\right|\mbox{\ ,}
\end{equation}
where $D=\delta_{c}(D(z)^{-1}-D(z_{0})^{-1})$ and $S=\sigma^{2}_{0}(M)-
\sigma^{2}_{0}(M_{0})$. Here, $\sigma^{2}_{0}(M)$ is the variance of the linear
power spectrum at $z=0$ smoothed with a ``top-hat'' filter of mass $M$,
$\delta_{c}=1.69$ is the usual threshold overdensity for spherical collapse,
and $D(z)$ is the growth
factor (Carroll, Press, \& Turner 1992).
In Figure 1, we show the number of subhalos at different redshifts $z$ that
end up as part of an $M_{0}$ system at $z_{0}$. In comparing the mass
functions of the DM subhalos after turnaround, $z_{ta}\simeq 24$, with
the observed one for old globular clusters (e.g., Harris \& Pudritz 1994),
one can see that the respective power-law slopes, ${\rm d}N/{\rm d}M\propto
M^{-1.8}$, agree quite well in the two cases. This fact suggests that
the DM subhalos assemble the requisite high-density gas
that might subsequently fragment to form the first globular clusters.

The gas can only fall into the DM subhalos if it is allowed to reach
temperatures below $\sim 10^{4}$K. This requirement explains the 
importance of the dwarf galaxy forming before the epoch of reionization.
After reionization, the intergalactic medium (IGM) is pervaded by
a general UV background that photoionizes the gas, and heats it to
temperatures $\ga 10^{4}$K (e.g., Barkana \& Loeb 2001).
Consequently, the gas in collapsing dwarf
systems would not `feel' the presence of the shallow DM potential
wells that are the suspected seeds for GC formation.

Recently, Weil \& Pudritz (2001) have investigated how the protogalactic
systems studied in this {\it Letter} would merge to form larger systems.
They do not, however, address what happens on scales smaller than $\sim$1 kpc,
the resolution limit of their simulation. Our work is complementary to
this larger scale study, as we focus on the star formation process in
the individual halos on sub-kpc scales.

\setcounter{figure}{0}
\begin{center} % fig.1
\epsfig{file=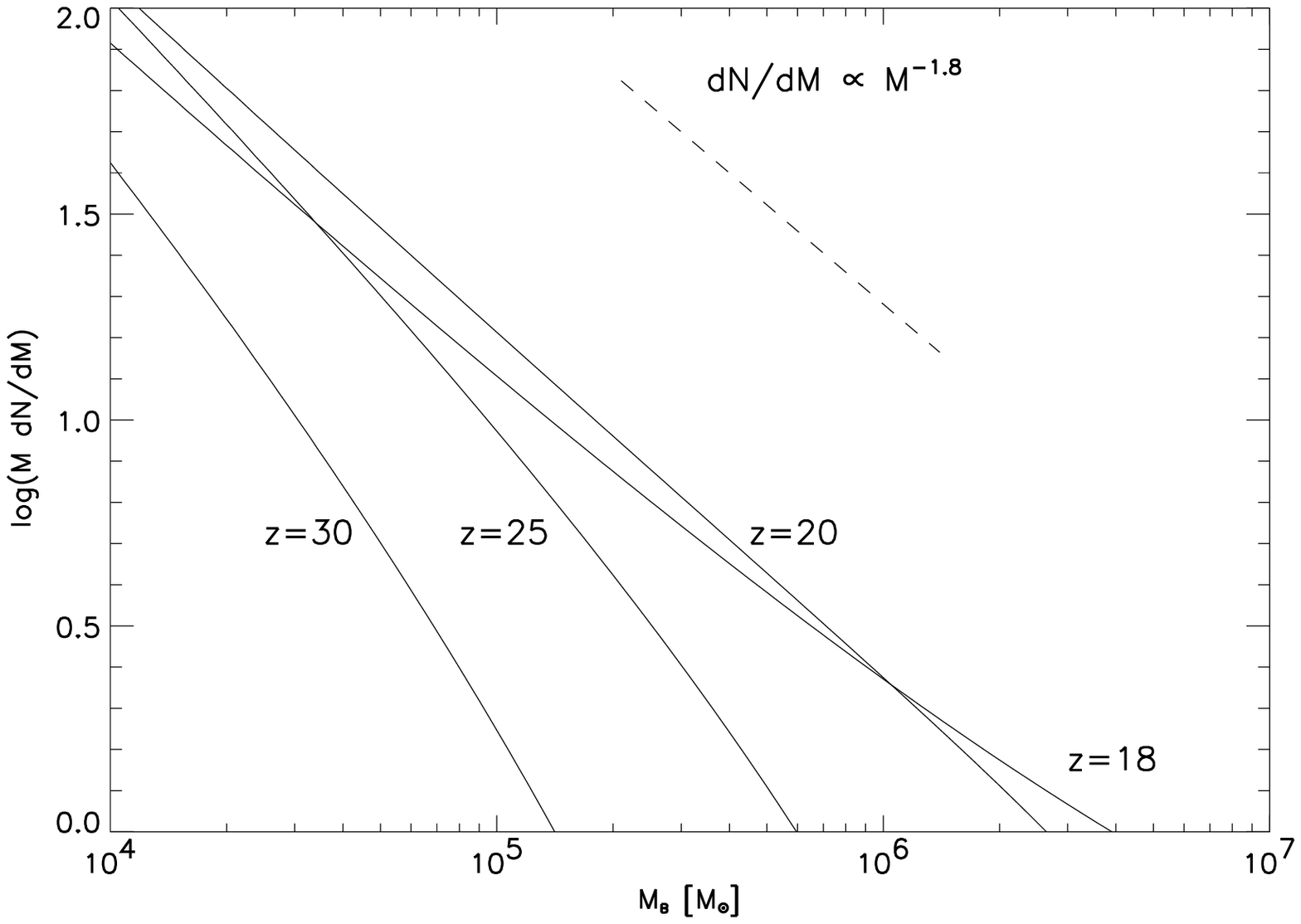,width=8.4cm,height=7.56cm}
\figcaption{Building up the dwarf galaxy. Average number of progenitor
halos vs. baryonic mass (in $M_{\odot}$),
formed at $z=$30, 25, 20, and 18 that, by the later time $z_{0}\simeq 15$,
will have merged into the dwarf galaxy halo of total mass $M_{0}=
2\times 10^{8}M_{\odot}$.
{\it Dashed line:} Observed mass function of old globular clusters.
It can be seen that, after the redshift of turnaround $z_{ta}\simeq 24$, 
the mass function of the DM subhalos is
very similar to the observed GC one.
\label{fig1}}
\end{center} % fig.1

\section{THE SIMULATIONS}
\subsection{Numerical Methodology}

The evolution of the dark matter and gas components is calculated with
a version of TREESPH (Hernquist \& Katz 1989), combining the smoothed
particle hydrodynamics (SPH) method with a tree gravity
solver.
To follow the thermal evolution of low metallicity
gas, we have implemented radiative cooling due to a trace amount of metals at
temperatures below $\sim 10^{4}$K, and due to atomic hydrogen and helium at
$T\ga 10^{4}$K (Bromm et al. 2001b).
Prior to reionization there are almost no ionizing photons, and we consequently
ignore heating due to photo-ionization.

We have devised an algorithm to merge SPH particles in high-density regions
in order to overcome the otherwise prohibitive time-step limitation, as
enforced by the Courant stability criterion. To follow the simulation
for a few dynamical times, we allow SPH particles to merge into more
massive ones, provided they exceed a predetermined density threshold,
$n_{th}\simeq 10^{3}$ cm$^{-3}$.
More details of the code are given in Bromm, Coppi, \& Larson (2001a).

%%\subsection{Initial Conditions}

Our simulation is initialized at $z_{i}=100$, by performing the following steps.
The collisionless DM particles are placed on
a cubical Cartesian grid, and are then perturbed 
by applying the 
Zeldovich approximation (see Bromm et al. 2001a).
%which also allows to self-consistently
%assign initial velocities. The power-law index is set to $n=-2.5$ which
The power-law index is set to $n=-2.5$ which
approximately describes the spectral behavior on the scale of dwarf galaxies.
%To fix the amplitude $A$, we specify the initial variance of the
%fluctuations
%%\begin{equation}
%$\sigma_{i}^{2}=A\sum k^{n}$.
%\mbox{\ \ \ .}
%\end{equation}
%The summation is over all contributing modes, where the minimum
%wavenumber is given by the overall size of the Cartesian box, and
%$k_{max}$ by the Nyquist frequency. Choosing $\sigma_{i}^{2}\simeq 0.04$,
%the rms fluctuation at the moment of collapse is
%$\sigma(z=15)\simeq 1$.
%This choice ensures that the substructure develops on a similar
%timescale as the overall collapse of the background medium.
Particles within a (proper) radius of $R_{i}=$ 950 pc 
are selected for the
simulation. The resulting number of DM particles is here $N_{\rm DM}=17074$.
Finally, the particles are set into rigid rotation and are endowed
with a uniform Hubble expansion.
\begin{center} % fig.2
\epsfig{file=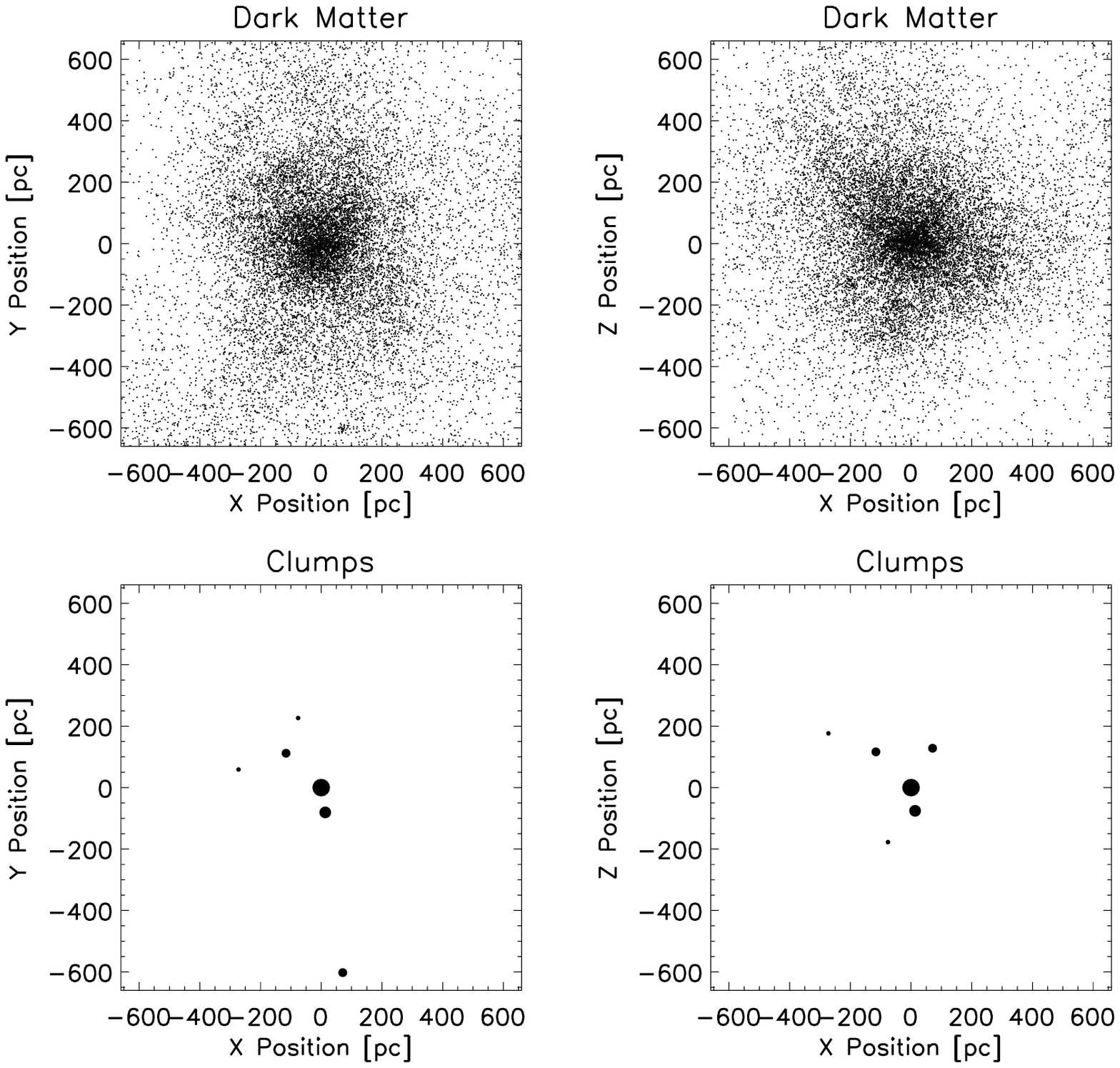,width=8.4cm,height=7.56cm}
\figcaption{
Morphology at $z\simeq 15$.
{\it Top row:} The DM component.
{\it Bottom row:} Distribution of clumps with dot size indicating clump
mass.
{\it Left panels:} Face-on view.
{\it Right panels:} Edge-on view.
The box size is 1.3 kpc.
The DM has undergone violent relaxation with the concurrent smoothing out
of the substructure. A small system of high-density clumps is left behind.
\label{fig2}}
\end{center} % fig.2

Angular momentum is added by assuming a spin-parameter
%\begin{equation}
%\lambda=\frac{L|E|^{1/2}}{G M^{5/2}}=0.05 \mbox{\ \ ,}
%\end{equation}
%$\lambda=L|E|^{1/2}/(G M^{5/2})=0.05$,
$\lambda=0.05$.
%where $L$, $E$, and $M$ are the total angular momentum, energy, and mass,
%respectively.
The collisional SPH particles ($N_{\rm SPH}=32768$) are randomly
placed to approximate
a uniform initial density, and 
are endowed with the same Hubble expansion and rigid rotation as the DM ones.
%For the initial gas temperature,
%we adopt:
%$T_{gas, i}\simeq 200$ K.

We assume that the gas is pre-enriched to a level of
$Z=10^{-2}Z_{\odot}$ as the result of previous star formation activity.
We have also carried out simulations with $Z=10^{-3}Z_{\odot}$ and
$Z=0$, but with the same DM realization, to investigate the effect of
varying the cooling physics.

\subsection{Results}

At first, the halo is still expanding, to break away from the Hubble
flow and to turn around at $z_{ta}\simeq 24$. At this point, the DM
has already developed a marked substructure in response to the initially
imprinted density fluctuations. The baryons have begun to fall into
the deepest DM potential wells. For this to happen, the gas has to be able
to cool below $\sim 10^{4}$K. Therefore, an efficient low-temperature
coolant has to be present: either metals at a level of $Z\ga 10^{-3}Z_{\odot}$
(see Bromm et al. 2001b) or molecular hydrogen. We have verified that the
DM induced assembly of gas into the subhalos leads to the same result, 
irrespective of whether the low-temperature cooling is provided by metals,
present at a level $10^{-3}$ or $10^{-2}Z_{\odot}$, or by H$_{2}$.
The assembly process, however, does not work if cooling is only due to lines
of atomic hydrogen. 
Once the gas cools and contracts to densities in excess of $n_{th}=10^{3}$
cm$^{-3}$, a sink particle is created. Initially, the sink particle has
a mass close to the resolution limit of the simulation, $M_{res}\sim 4\times
10^{4}M_{\odot}$. Subsequently, a sink particle grows in mass by accreting
surrounding, diffuse gas, and by merging with other sink particles.

In Figure 2, the situation towards the end of the virialization
process is shown.

\begin{center} % fig.3
\epsfig{file=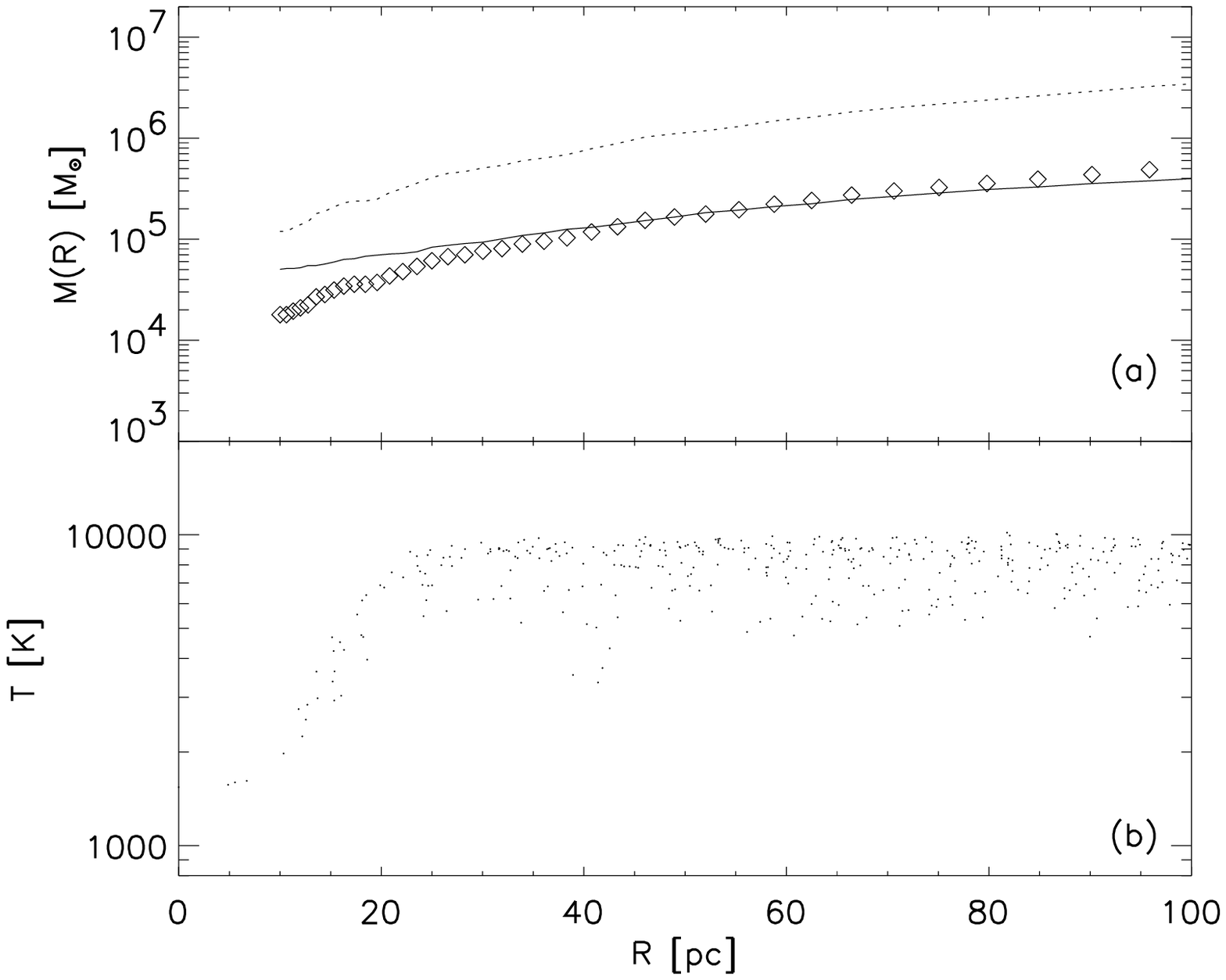,width=8.4cm,height=7.56cm}
\figcaption{Surroundings of the first high-density clump at $z\simeq 24$.
({\it a}) Enclosed mass (in $M_{\odot}$) vs. radial distance (in pc)
from the density maximum.
{\it Dashed line:} Mass profile of the dark matter.
{\it Solid line:} Mass profile of the baryons.
{\it Diamond-shaped symbols:} Baryonic mass profile scaled from the DM
profile with the primordial ratio of gas to DM mass ($\sim 0.18$).
({\it b}) Gas temperature (in K) vs. radial distance.
It can be seen that the drop in gas temperature at $R\simeq 20$ pc 
approximately coincides with the baryonic mass profile beginning to
deviate from the primordial scaling (see ({\it a})).
\label{fig3}}
\end{center} % fig.3

The collapse has resulted in the formation of 6 high density clumps, with
masses (in units of $10^{5}M_{\odot}$):
0.4, 0.5, 2, 3,
13, and 220.
The central clump is untypically massive for a GC. This most massive
clump, however, differs from the other ones in that it has been assembled
over a long period of time ($\Delta t\sim 10^{8}$yr), and is therefore
expected to be chemically inhomogeneous and possibly subject to vigorous
negative feedback due to multiple episodes of star formation. 
The lower mass limit, $M_{lower}$, is set by the resolution limit of
the simulation, $M_{res}\sim 5\times 10^{4} M_{\odot}$. There is, however,
also a physical reason for a lower mass cutoff.
During the collapse of the halo
the Jeans mass obeys
the relation: $M_{J}\ga 10^{5}M_{\odot}$
(see Fig. 4), and gas pressure will therefore prevent the baryons from
falling into the smallest DM subhalos.

The gaseous clumps appear to have lost their individual DM halos and
are embedded in a general, rather smooth DM distribution.
This almost
complete erasure of the DM substructure might be due to insufficient
numerical resolution, as has been recently demonstrated on the scale
of large (Milky-Way size) galaxies, as well as of clusters of galaxies
(e.g., Moore et al. 1999). 
The small-scale regime explored here, however, might be special in that
the slope of the mass variance becomes almost flat. All scales collapse
virtually at the same time, and the individual DM subhalos consequently 
experience exceptionally strong tidal fields. Only in the smallest galaxies,
therefore, might the dissipatively condensed gas clumps lose the association
with the DM halos that had given birth to them.
Future high resolution simulations will
test whether this is indeed the case.
Our model could provide an explanation for the existence of
co-moving groups of GCs in the halos of large galaxies, as claimed
by Ashman \& Bird (1993) in the case of M31.

In Figure 3, we show the properties of the gas in the vicinity of the first
high-density clump. At $r\sim 20$ pc, the gas mass profile begins to
deviate from the primordial behavior, $M_{\rm B}(r)=0.18 M_{\rm DM}(r)$.

\begin{center} % fig.4
\epsfig{file=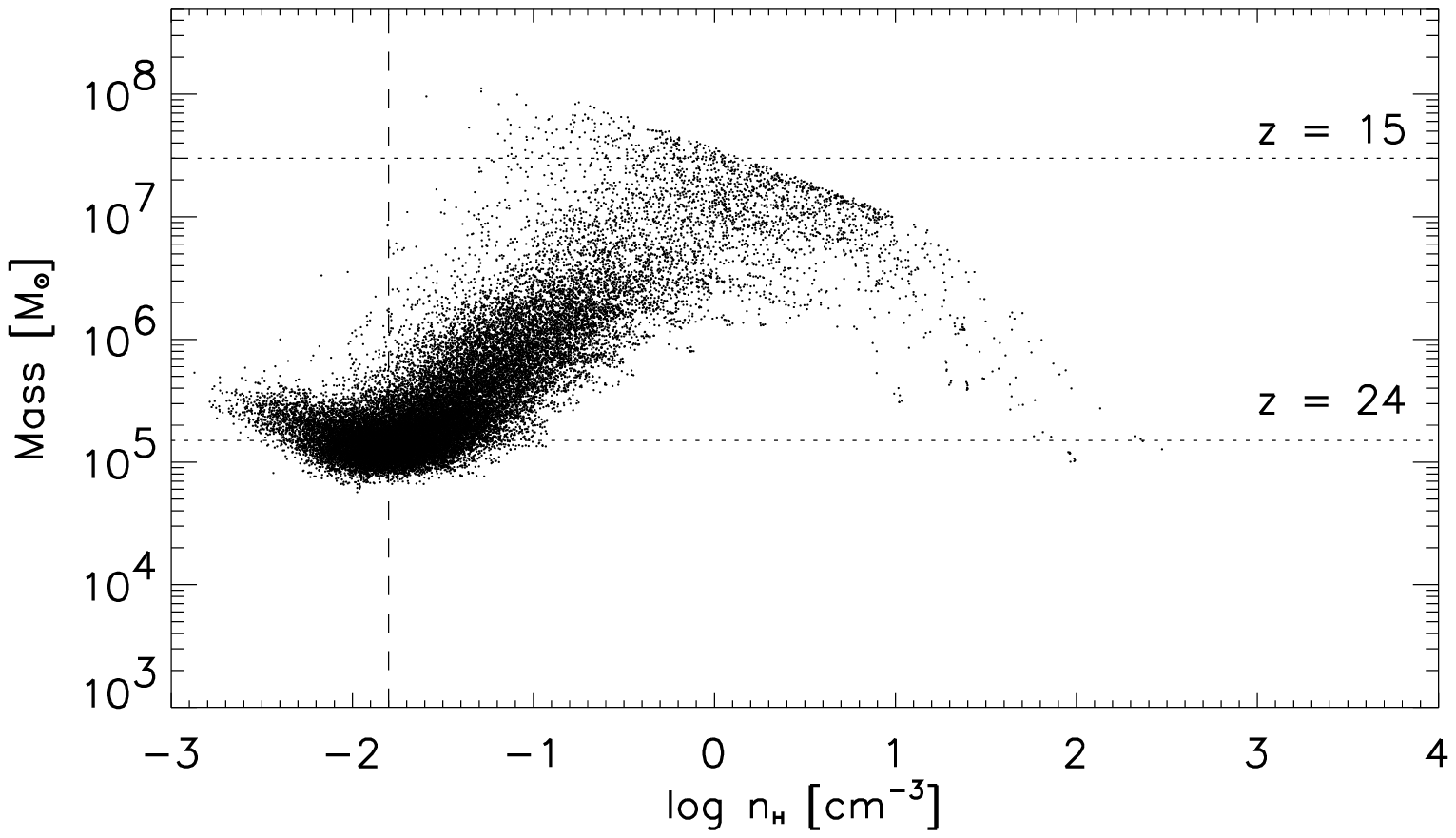,width=8.4cm,height=7.56cm}
\figcaption{The bottom-up assembly of gas in a collapsing dwarf galaxy.
Gas mass (in $M_{\odot}$) vs. hydrogen number density (in cm$^{-3}$).
{\it Small dots:} Jeans mass evaluated for every SPH particle at the
moment of turnaround, $z_{ta}\simeq 24$.
{\it Dotted lines:} Baryonic mass of a halo corresponding to a 3$\sigma$
peak at redshifts $z=24$, and 15.
{\it Vertical dashed line:} Condition that the free-fall time equals
the Hubble time at $z=15$. Only the particles to the right of this line,
comprising $\sim 70$\% of the total gas mass, have a chance to be incorporated
into a high-density clump.
\label{fig4}}
\end{center} % fig.4

Closer in, at $r\sim 10$ pc, the gas becomes self-gravitating.
The DM density in the virialized subhalo is $\rho_{\rm DM}\sim
10^{-21}$g cm$^{-3}$, corresponding to a gas density of
$n \sim 10^{2}$cm$^{-3}$. Since sink particles are created at
$n_{th}=10^{3}$cm$^{-3}$, the high-density clumps physically
represent self-gravitating clouds. These clouds are the possible
progenitors of globular clusters. Notice the importance of the gas being
able to cool below $\sim 10^{4}$K, in order to fall into the DM potential
well (Fig. 3b).
A radius of $\sim 10$ pc constitutes an upper limit to the true
size of a GC. The final cluster is expected to be more compact than
the simulated clumps. GCs are observed to show a lack of correlation
between mass and radius (e.g., Ashman \& Zepf 2001). Any mass-radius
relation in our simulation would be a numerical artefact, since sink
particles (clumps) are created at a fixed density. Determining the
precise nature of the $M-R$ relation and the true radial extent of
the GCs require higher-resolution simulations which we plan to carry out 
in future work.

In Figure 4, we estimate the mass range of the resulting
clumps. We compute the Jeans mass for each gas particle using the total
(DM $+$ baryonic) density at a redshift of 24. We also show (horizontal lines)
the baryonic mass of a 3$\sigma$ peak at redshift 24 (turnaround) and
15 (virialization).
At first the highest density gas collapses into a
$\sim 10^{5}M_{\odot}$ halo (see Fig. 3).
Subsequently, more massive clumps form,
up to the collapse of the dwarf galaxy itself.
For gas with $T\ga 10^{4}$K, 
characteristic of photoionization-heated gas, the Jeans mass would be
$M_{J}\ga 10^{7}M_{\odot}$, hence making it impossible to form
any high density clumps in the small DM subhalos. Again, this shows that
the GC formation mechanism proposed in this {\it Letter} can only
operate at redshifts beyond reionization.

\section{SUMMARY AND CONCLUSIONS}

%We have presented an efficient mechanism to assemble large amounts of gas
%into high density clouds, operating in dwarf galaxy sized halos at redshifts
%beyond the epoch of reionization. The collapse and virialization of the
The collapse and virialization of the 
dwarf galaxy results in the formation of six high density clumps, with 
masses $10^{5}\la M \la 10^{7}M_{\odot}$, and typical radii $\sim 10$ pc.
The resulting clump masses are determined by the mass spectrum of the
DM substructure, provided the gas can cool sufficiently to fall into
the shallow DM potential wells. This is the case in the presence of an
efficient coolant that is able to operate at $T\la 10^{4}$K, but is prohibited
after the universe underwent reionization.
In our simulations, the highly condensed gas clouds have lost their
individual DM subhalos, which we tentatively ascribe to the
exceptionally strong tidal forces acting during the relaxation of a
dwarf galaxy at high redshift.
%%The systems investigated
%%in this paper might be candidates for the fragments proposed by Searle \& Zinn
%%(1978) as the basic building blocks that accreted onto the incipient Galaxy
%%and contributed at least a fraction of its globular clusters.

%An attractive feature of the proposed mechanism is that it can naturally account
%for the observed GC central velocity dispersion of $\sim 10$ km s$^{-1}$.
%Since the virial temperature of the DM subhalos is similar to that
%of the parent halo, $T_{vir}\sim 10^{4}$K, the velocity dispersion at the
%time of virialization has the correct order of magnitude. Recent stellar
%dynamical simulations have shown that two-body relaxation does not significantly
%alter the overall value of 
%the velocity dispersion imprinted initially (Giersz \& Heggie 1994).

The expected metallicity distribution of clusters formed in this way
needs further investigation. We note that the time between turnaround
and virialization, which marks the interval over which GCs
are assembled in our model, is around $10^{8}$yr, and thus
in principle there is time for the nucleosynthetic products from one
cluster to reach the gas destined to form another cluster. It is
currently unclear whether this could explain the significant spread 
in metallicity observed among the clusters in the Fornax and
Sagittarius dwarfs 
(e.g., Buonanno et al. 1998), or whether these data require that
some of the clusters form at significantly later epochs.

\acknowledgments{We would like to thank Richard Larson, Lars Hernquist,
and the anonymous referee for helpful comments. We are indebted to
Lars Hernquist for making available to us a version
of TREESPH, and to Andrea Ferrara for providing us with his low-metallicity
cooling functions.
This work has been supported by the
``European Community's Research Training Network under contract
HPRN-CT-2000-0155, Young Stellar Clusters.''
%EC RTN network ``The Formation and Evolution of Young Stellar Clusters''.
}

%\vfill\eject

\clearpage


\begin{references}
\reference{}Ashman, K. M. 1990, \mnras, 247, 662
\reference{}Ashman, K. M., \& Bird, C. M. 1993, \aj, 106, 2281
\reference{}Ashman, K. M., \& Zepf, S. E. 1998, 
Globular Cluster Systems (Cambridge: Cambridge Univ. Press)
\reference{}Ashman, K. M., \& Zepf, S. E. 2001, \aj, 122, 1888 
\reference{}Barkana, R., \& Loeb, A. 2001, Physics Reports, 349, 125
\reference{}Blumenthal, G.R., Faber, S.M., Primack, J.R., \& Rees, M.J. 1984,
Nature, 311, 517
%\reference{}Becker, R. H., et al. 2001, \aj, in press
%(astro-ph/0108097)
%%\reference{}Bower, R. G. 1991, \mnras, 248, 332
\reference{}Bromm, V., Coppi, P. S., \& Larson, R. B. 2001a, \apj, in press
%\reference{}Bromm, V., Coppi, P. S., \& Larson, R. B. 2001a, \apj, 564, 1 
(astro-ph/0102503)
\reference{}Bromm, V., Ferrara, A., Coppi, P. S., \& Larson, R. B. 2001b,
\mnras, 328, 969
\reference{}Buonanno, R., Corsi, C. E., Zinn, R., Fusi Pecci, F., Hardy, E., \& Suntzeff, N. B.
1998, \apj, 501, L33
\reference{}Carroll, S.M., Press, W.H., \& Turner, E. L. 1992, \araa, 30, 499
\reference{}Cen, R. 2001, \apj, 560, 592
\reference{}C\^{o}t\'{e}, P., West, M. J., \& Marzke, R. O. 2001, \apj, in press
(astro-ph/0111388)
\reference{}Fall, S. M., \& Rees, M. J. 1985, \apj, 298, 18
%\reference{}Giersz, M., \& Heggie, D. C. 1994, \mnras, 268, 257
\reference{}Gnedin, N. Y. 2000, \apj, 535, 530
\reference{}Harris, W. E., \& Pudritz, R. E. 1994, \apj, 429, 177
\reference{}Hernquist, L., \& Katz, N. 1989, \apjs, 70, 419
\reference{}Kang, H., Shapiro, P.R., Fall, S. M., \& Rees, M. J. 1990,
\apj, 363, 488
\reference{}Lacey, C., \& Cole, S. 1993, \mnras, 262, 627
\reference{}McLaughlin, D. E., \& Pudritz, R. E. 1996, \apj, 457, 578
\reference{}Moore, B. 1996, \apj, 461, L13
\reference{}Moore, B., Ghigna, S., Governato, F., Lake, G., Quinn, T.,
Stadel, J., \& Tozzi, P. 1999, \apj, 524, L19
\reference{}Murray, S. D., \& Lin, D. N. C. 1992, \apj, 400, 265
\reference{}Nakasato, N., Mori, M., \& Nomoto, K. 2000, \apj, 535, 776
\reference{}Padoan, P., Jimenez, R., \& Jones, B. 1997, \mnras, 285, 711
\reference{}Peebles, P. J. E., \& Dicke, R. H. 1968, \apj, 154, 891
\reference{}Peebles, P. J. E. 1984, \apj, 277, 470
%\reference{}Rees, M. J. 2000, New Perspectives in Astrophysical Cosmology
%(Cambridge: Cambridge Univ. Press), 36
\reference{}Rosenblatt, E.I., Faber, S.M., \& Blumenthal, G.R. 1988,
\apj, 330, 191
%%\reference{}Searle, L., \& Zinn, R. 1978, \apj, 225, 357
%\reference{}Tegmark, M., Silk, J., Rees, M. J., Blanchard, A., Abel, T., \& Palla, F.
%1997, \apj, 474, 1
\reference{}Weil, M. L., \& Pudritz, R. E. 2001, \apj, 556, 164
\reference{}Whitmore, B. C., \& Schweizer, F. 1995, \aj, 109, 960
%\reference{}Zel'dovich, Y. B. 1970, A\&A, 5, 84


\end{references}
\end{document}